\documentclass[showpacs,twocolumn,showkeys,amsmath,amssymb,pra,superscriptaddress]{revtex4-1}

\usepackage{graphicx}

\newcommand{\f}{a}
\newcommand{\xx}{h}

\begin{document}
\title{Fractional photon-assisted tunnelling of ultra-cold atoms in periodically shaken double-well lattices} 

\author{Martin Esmann}
\affiliation{Physics Department, Harvard University, Cambridge, MA 02138, USA}
\affiliation{Institut f\"ur Physik, Carl von Ossietzky Universit\"at, D-26111 Oldenburg, Germany}

\author{Jonathan D. Pritchard}
\affiliation{Department of Physics, University of Strathclyde, Glasgow, G4 0NG,  United Kingdom}

\author{Christoph Weiss}
\affiliation{Institut f\"ur Physik, Carl von Ossietzky Universit\"at, 26111 Oldenburg, Germany}
\affiliation{Department of Physics, Durham University, Durham DH1 3LE, United Kingdom}
\email{Christoph.Weiss@durham.ac.uk}

\date{13 September 2011}

\begin{abstract}
Fractional photon-assisted tunnelling is investigated both numerically and analytically in a double-well lattice.  While integer photon-assisted tunnelling is a single-particle effect, fractional photon-assisted tunnelling is an interaction-induced many-body effect. Double-well lattices with few particles in each double well are ideal to study this effect far from the mean-field effects. It is predicted that the 1/4-resonance is observable in such systems. Fractional photon-assisted tunnelling provides a physically relevant model for which $N$-th order time-dependent perturbation theory can be large although all previous orders are small.
\end{abstract}



\maketitle 


\section{Introduction}

\begin{figure}
\includegraphics[clip,width=.7\columnwidth]{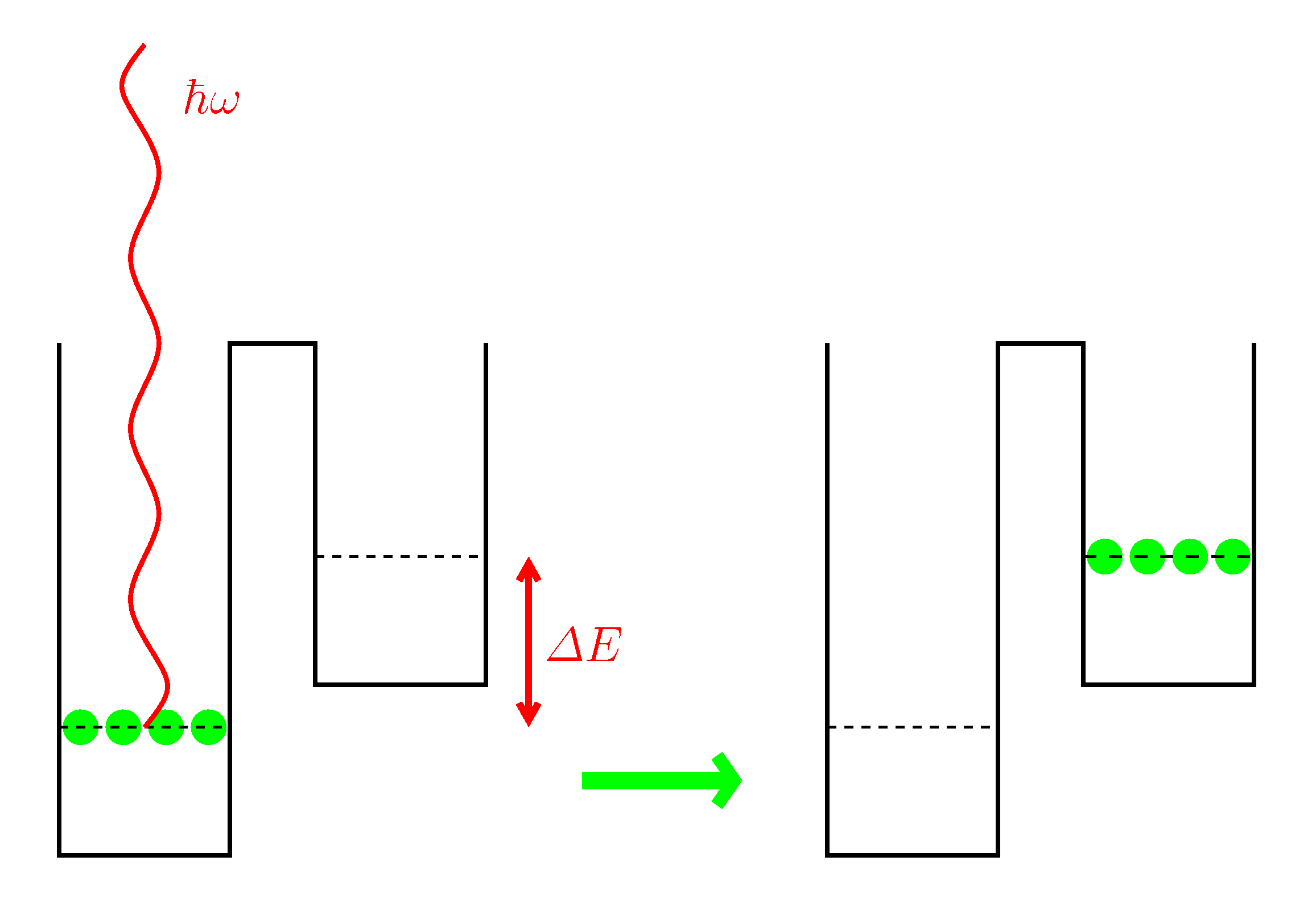}
\caption{\label{fig:title}(Colour online) Sketch of 1/4-`photon' resonance: One photon has
enough energy to make four particles tunnel ($\hbar\omega = 4\Delta E$). The ``photons'' are time-periodic potential modulations in
the kilo-Hertz regime.
}
\end{figure}

Optical lattices are an important system for research with ultra-cold atoms~\cite{LewensteinEtAl07,BlochEtAl08,Yukalov09}. 
Experimental developments enable the creation of lattices of controllable double-well potentials~\cite{SebbyStrableyEtAl07,YukalovYukalova09} that can be engineered such that the tunnelling between neighbouring double wells can be discarded~\cite{Foelling10}, allowing treatment of the system as a single double well. Loading the lattice from a Mott-insulator state allows deterministic population of fewer than six atoms in each well~\cite{Folling07,CheinetEtAl08}. Combined with the ability to count the atoms in each well, this makes the double well system ideal for investigating fractional photon-assisted tunnelling via periodic shaking of the lattice that is typically a small effect in other systems~\cite{EckardtEtAl05,TeichmannEtAl09,XieEtAl10}.  The ``photons'' are  time-dependent potential modulations in the kilo-Hertz regime; a sketch of the 1/4-photon resonance can be seen in Fig.~\ref{fig:title}.

Research on periodic shaking has focused on effects ranging from destruction of tunnelling~\cite{GrossmannEtAl91,KierigEtAl08,GongEtAl09}  over tunnelling-control~\cite{Holthaus92,Weiss06b,LuEtAl10}, field-induced  barrier transparency~\cite{Longhi10}, two-dimensional solitons~\cite{SudheeshEtAl10}, super Bloch oscillations~\cite{HallerEtAl10,KudoMonteiro11,ArlinghausHolthaus11}, phase-jumps~\cite{RidingerWeiss09,ClearyEtAl10} and dynamics of bound pairs in optical lattices~\cite{KudoEtAl09,WeissBreuer09} or NOON-states~\cite{StieblerEtAl11}. Complementary studies of transport of ultra-cold atoms in lattices include controlled transport of Bose-Einstein condensates (BECs) between two wells~\cite{NesterenkoEtAl10}. Experiments include control of superexchange interactions~\cite{TrotzkyEtAl08,ChenEtAl11} and directed transport via a Hamiltonian quantum ratchet~\cite{SalgerEtAl09}.

While integer photon-assisted tunnelling~\cite{KohlerSols03,SiasEtAl08} is essentially a single particle effect that survives interactions, fractional photon-assisted tunnelling is a true many-particle effect which only occurs for interacting particles. Rather than being the small effect predicted in Refs.~\cite{EckardtEtAl05,TeichmannEtAl09,XieEtAl10},  fractional photon-assisted tunnelling like the 1/2-photon resonance can be a large effect for two particles per double well~\cite{EsmannEtAl11,ChenEtAl11}. A related experiment on photon-assisted tunnelling of strongly correlated atoms can be found in Ref.~\cite{MaEtAl11}.

In this Letter we investigate fractional-photon assisted tunnelling of ultra-cold atoms in a double well potential. The paper is organised as follows: Section~\ref{sec:model} introduces the model describing atoms in the periodically shaken lattice, whilst Sec.~\ref{sec:BEC} explores the effect for both few and hundreds of atoms per well. In Sec.~\ref{sec:4} we demonstrate that for four particles, both the 1/4- and the 1/2-resonance will provide clear experimental signatures. Section~\ref{sec:avoided} explains a feature in the photon-assisted tunnelling plot reminiscent of avoided crossings. In Sec.~\ref{sec:N} we show that for some parameters, even though the first few orders of time-dependent perturbation theory might be small, higher order perturbation theory can still correctly predict the fact that fractional photon-assisted tunnelling is a large effect. 

\section{\label{sec:model}Model}

\subsection{Hamiltonian}
For both the case of a few ultra-cold atoms, or a small BEC loaded into the double-well potential modulated at frequency~$\omega$, the system can be described using a many-body Hamiltonian with a two-mode approximation as follows~\cite{CheinetEtAl08,MilburnEtAl97,LipkinEtAl65}
\begin{eqnarray}
\label{eq:H}
\hat{H} &=& -\frac{\hbar\Omega}2\left(\hat{c}_1^{\dag}\hat{c}_2^{\phantom\dag}+\hat{c}_2^{\dag}\hat{c}_1^{\phantom\dag} \right) + \hbar\kappa\left(\hat{c}_1^{\dag}\hat{c}_1^{\dag}\hat{c}_1^{\phantom\dag}\hat{c}_1^{\phantom\dag}+\hat{c}_2^{\dag}\hat{c}_2^{\dag}\hat{c}_2^{\phantom\dag}\hat{c}_2^{\phantom\dag}\right)\nonumber\\
&+&\hbar\big(\mu_0+\mu_1\sin(\omega t)\big)\left(\hat{c}_2^{\dag}\hat{c}_2^{\phantom\dag}-\hat{c}_1^{\dag}\hat{c}_1^{\phantom\dag}\right)\;.
\end{eqnarray}
The operators $\hat{c}^{{\phantom{\dag}}}_j$/$\hat{c}^{\dag}_j$ annihilate/create a boson in well~$j$;
$\hbar\Omega$ is the tunnelling splitting, $\hbar\mu_0$ denotes
the tilt between well~1 and well~2 and $\hbar\mu_1$ is the driving amplitude (cf.\ Fig.~\ref{fig:title}). The on-site pair interaction is denoted by $2\hbar\kappa$. For calculations beyond this model see,  e.g., Refs.~\cite{TrimbornEtAl09,SakmannEtAl09,ZollnerEtAl08,GrondEtAl11}. Here, we use the two-mode approximation as the experiment~\cite{ChenEtAl11} demonstrates that this approximation describes the physics of the 1/2-photon resonance. Our focus lies in identifying and understanding interesting signatures of photon-assisted tunnelling rather than quantitative predictions;calculations including effects of higher energy levels (cf.\ Ref.~\cite{ChenEtAl11}) will subsequently depend on precise experimental details like the depth of the lattice.

In order to characterise the photon-assisted tunnelling, we use the experimentally measurable time-averaged particle transfer probability, 
\begin{equation}
\label{eq:transfer}
\left\langle P_{\rm trans}\right\rangle_T=\frac{1}{NT}\int_0^T{\left\langle \Psi\left(t\right)\left|\hat{c}_2^{\dag}\hat{c}_2^{\phantom\dag}\right|\Psi\left(t\right)\right\rangle}dt\;.
\end{equation}

Rather than trying to understand the physics by directly solving the Schr\"odinger equation corresponding to the Hamiltonian~(\ref{eq:H}), we derive an equivalent set of differential equations~\cite{EsmannEtAl11}.
We choose the Fock basis
$
   |\nu\rangle \equiv |N-\nu,\nu\rangle 
$   
for which the label   
$   
   \nu=0\ldots N
$
refers to a state with $N-\nu$~particles in well~$1$,
and $\nu$~particles in well~$2$. In this basis, the Hamiltonian~(\ref{eq:H}) can be written as the 
sum of two $(N+1)\times(N+1)$-matrices,
\begin{equation}
\label{eq:Hsum}
   H = H_0(t) + H_1 \; ,
\end{equation}
where the diagonal matrix~$H_0(t)$ includes both the interaction between 
the particles and the applied potential difference 
while the non-diagonal matrix~$H_1$ contains the tunnelling-terms of Eq.~(\ref{eq:H}).

The ansatz
\begin{equation}
   \langle \nu|\psi(t)\rangle = 
   \f_\nu(t)\exp\left[-\frac{{i}}{\hbar}\int
   \langle \nu | H_0(t)|\nu\rangle d t\right],
\end{equation}
which is based on the interaction picture, turned out to be useful to solve the Schr\"odinger equation~\cite{TeichmannEtAl09}. The Schr\"odinger equation  then is equivalent to the set of $N+1$ differential equations:
\begin{align}
\label{eq:equiv}
   &{i}\hbar\dot{\f}_{\nu}(t) = 
   \langle\nu | H_1 |\nu\!+\!1\rangle\xx_{\nu}(t){\f}_{\nu+1}(t)\quad
\\\nonumber
   &+  \langle\nu|H_1|\nu\!-\!1\rangle \xx_{\nu-1}(t)^*{\f}_{\nu-1}(t) ,\quad \nu =0,1,\ldots N
\end{align}
which uses the notation
$
   a_{-1}(t) \equiv a_{N+1}(t) \equiv 0
$;
the phase factors read:
\begin{eqnarray}
\label{eq:phase}
   \xx_{\nu}(t) =\hspace*{6cm}  \\ \nonumber
\exp\left(
{\textstyle {i}\left[2(N - 1 - 2\nu)\kappa t 
   -2\mu_0t+2\mu_1\cos(\omega t)/\omega\right] }
\right).
\end{eqnarray}

\subsection{\label{sec:timedependent}Time-dependent perturbation theory}
In the following, we always use the experimentally realistic initial condition that all the atoms are in the lower well at $t=0$, i.e.\ $a_0(0)=1$~\cite{Folling07,CheinetEtAl08}. Then at a later time $t$, zeroth-order time-dependent perturbation theory gives the Fock-state amplitudes as:
\begin{align}
 a_0^{(0)}(t)&\equiv 1\\
 a_j^{(0)}(t)&\equiv 0,\quad j>0 \;.
\end{align}

The first non-zero order of $a_j^{(k)}$ is obtained for $k=j$:
\begin{align}
\label{eq:leading}
a_j^{(j)}(t) \equiv& i\frac{\sqrt{(N-j+1)j}}2\Omega\nonumber\\\times&\int_0^{t}d{\tau}\,h_{j-1}({\tau})^*a_{j-1}^{(j-1)}({\tau}),\quad j\ge 1,
\end{align}
where $^*$ denotes the complex conjugate.

\subsection{\label{sec:understanding}Tunnelling dynamics}
To understand the tunnelling dynamics, we expand the oscillatory term at frequency $\omega$ in the phase factors of Eq.~(\ref{eq:leading}) in terms of Bessel functions~\cite{Abramowitz84}
\begin{equation}
\label{eq:bessel}
e^{iz\cos(\omega t)}=\sum_{k=-\infty}^{\infty}J_{k}(z)i^ke^{ik\omega t}\,.
\end{equation}
Including all these terms in analytic calculations does, in principle, lead to analytic results for the tunnelling. However, evaluating these analytic formulae is numerically much more intensive than solving the time-dependent Schr\"odinger equation~\cite{TeichmannEtAl09}. Combining the rotating-wave-approximation based approach which includes only the slowly oscillating terms of $h_j^*$~\cite{EsmannEtAl11} with the above time-dependent perturbation theory leads to a simpler form for the non-zero perturbations of Eq.~(\ref{eq:leading}):
\begin{align}
\label{eq:leading2}
a_j^{(j)}(t) \equiv& i^{1-k_j}\frac{\sqrt{(N-j+1)j}}2\Omega J_{k_j}\left({\textstyle \frac{2\mu_1}{\omega}}\right)\nonumber\\\times&\int_0^{t}d{\tau}\exp(i\eta^{(j)}_{k_j}{\tau})a_{j-1}^{(j-1)}({\tau}),\quad j\ge 1
\end{align}
where the integer $k_j$ is chosen such that $|\eta_{k}|$, with
\begin{align}
\label{eq:eta}
\eta_{k}^{(j)}\equiv-k\omega+2\mu_0-2[N-(2j-1)]\kappa\;, \\  j=1\ldots N\;.\nonumber
\end{align}
is minimised at $k=k_j$. As in Ref.~\cite{EsmannEtAl11}, it might sometimes be preferable to minimise $|\sum_{\nu}^j\eta^{(\nu)}_{k_{\nu}}|$ rather than each $\eta^{(j)}_{k}$ separately.  For other cases (e.g.\ near a zero of one of the Bessel functions) more than one term will have to be included in the above sums. We define the number of ``photons'' involved in the tunnelling process,
\begin{equation}
\label{eq:nphoton}
\#_{\rm photons} = \sum_{j=1}^N k_j,
\end{equation}
such that it corresponds to the total ``energy'' transferred (rather than taking the sum of the moduli).

\begin{figure}
\includegraphics[width=\linewidth]{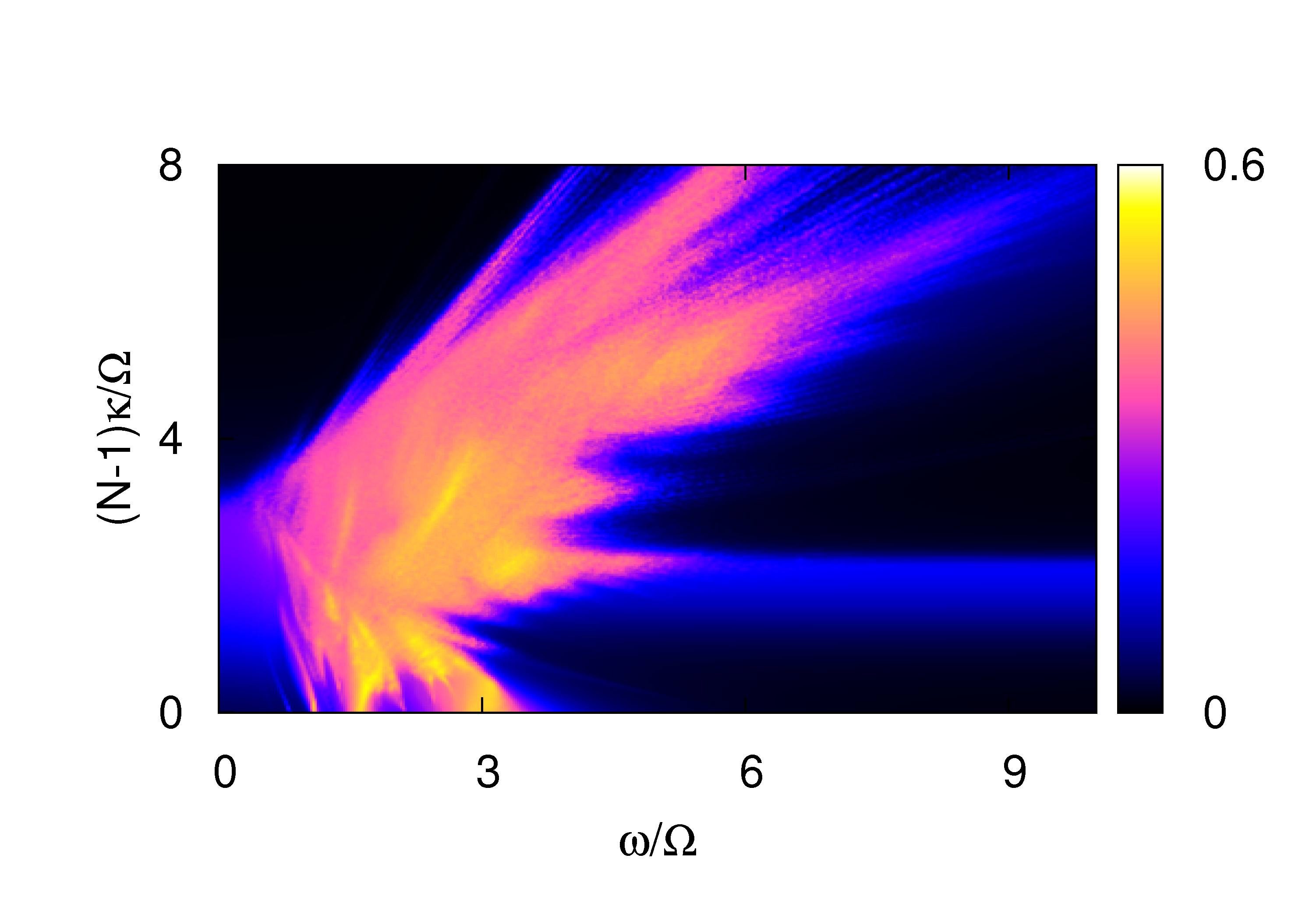}
\caption[test]{\label{fig:100}(Colour online) Two-dimensional projection of the time-averaged particle transfer~(\ref{eq:transfer}) as a function of both driving frequency $\omega$ and interaction $\kappa$ for  $N=100$ particles. The averaging time is $\Omega T=100$, the driving amplitude $2\mu_1/\omega = 1.8$ and the tilt $\mu_0 =1.5\Omega$. The one-photon resonance is clearly visible: it starts at the point defined by $\kappa=0$ and $\omega =3\Omega$ and moves towards lower frequencies for larger interactions. While many-integer-photon-resonances like the two-photon resonance near $\kappa=0$ and $\omega =1.5\Omega$ are also visible, fractional photon-assisted tunnelling like the one-half resonance near $\omega =6\Omega$ (cf.\ Ref.~\cite{EckardtEtAl05}) are not visible on this scale because they are only a small effect for large particle numbers.}
\end{figure}

\section{\label{sec:BEC}BEC or few particles per double well?}
In order to demonstrate why it is necessary to use a few atoms per double well as opposed to a BEC to observe fractional resonances, we compare the results obtained for $N=100$ particles with $N=4$. Figure~\ref{fig:100} shows the time averaged particle transfer probability for $N=100$ atoms initially in the lower well. While there are no resonances visible at higher frequencies, the one-photon-resonance which starts at $\omega=3\Omega$ for small interactions is clearly visible. For a BEC initially in the upper well, the one-photon resonance would move toward higher frequencies for increasing interaction.  

\begin{figure}
\includegraphics[width=\linewidth]{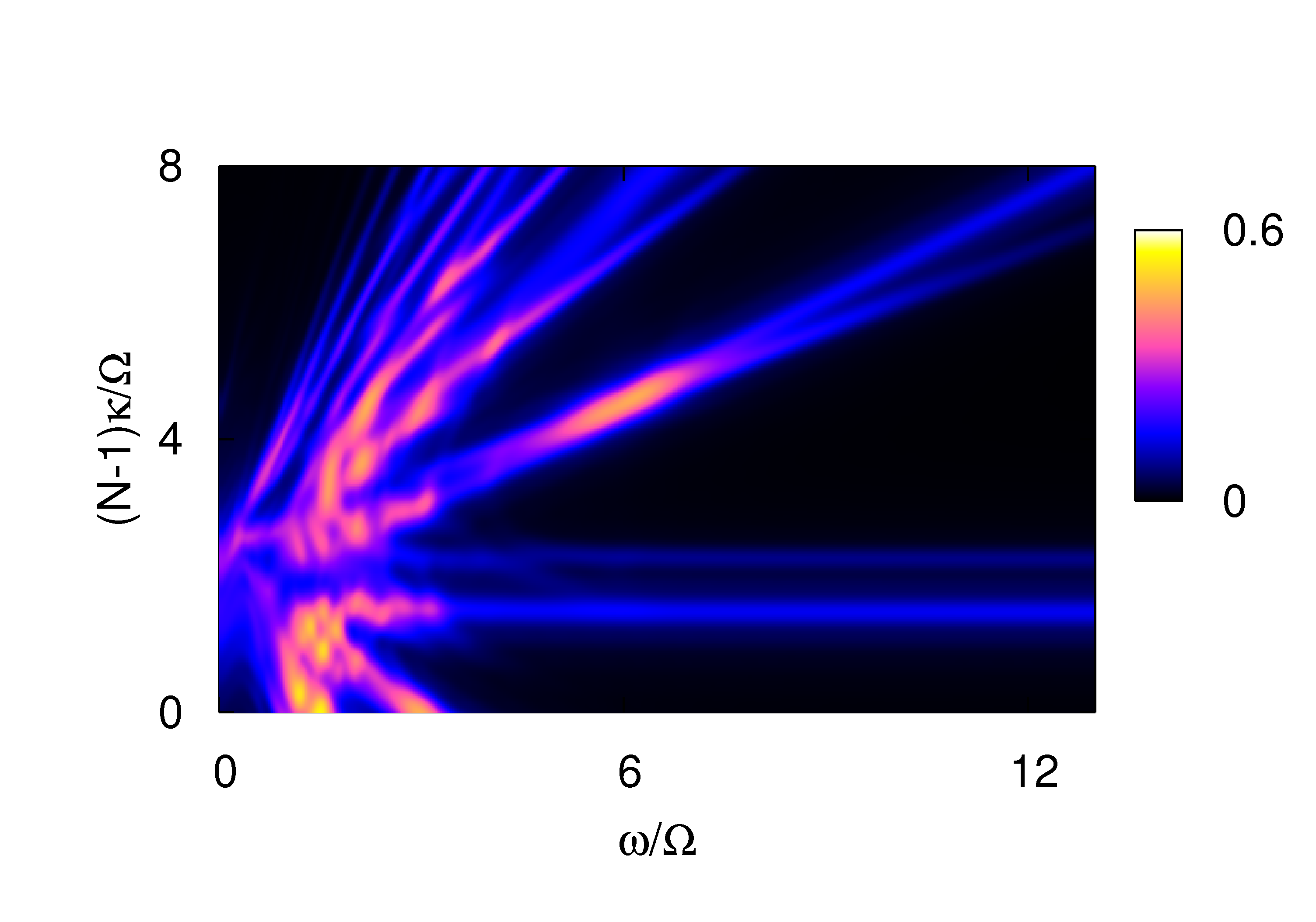}
\caption{\label{fig:4_10}(Colour online) Time-averaged transfer probability to well~2 for $N=4$ calculated using the same parameters as in Fig.~\ref{fig:100} but with an averaging time of $\Omega T =10$. Experimentally, this could be realised in the double-well lattice of Refs.~\cite{Folling07,CheinetEtAl08}. }
\end{figure}

Figure~\ref{fig:4_10} shows the time-averaged transfer probability for N=4 calculated using the same parameters as Fig.~\ref{fig:100} with an averaging time of $\Omega T = 10$. For this shorter averaging time-scale, Fig.~\ref{fig:4_10} displays many features also seen in Fig.~\ref{fig:100}. The visible lines are either tunnelling resonances which can be understood on the single-particle level like the 1-photon resonance near $\kappa\approx 0$ and $\omega\approx 3\Omega$ or the horizontal lines. These correspond to, e.g.,  the energy of all particles being in the lower well and one particle having tunnelled being equal. The straight lines with non-zero gradient correspond to adding one or several photons to those horizontal lines.

\begin{figure}
\includegraphics[width=\linewidth]{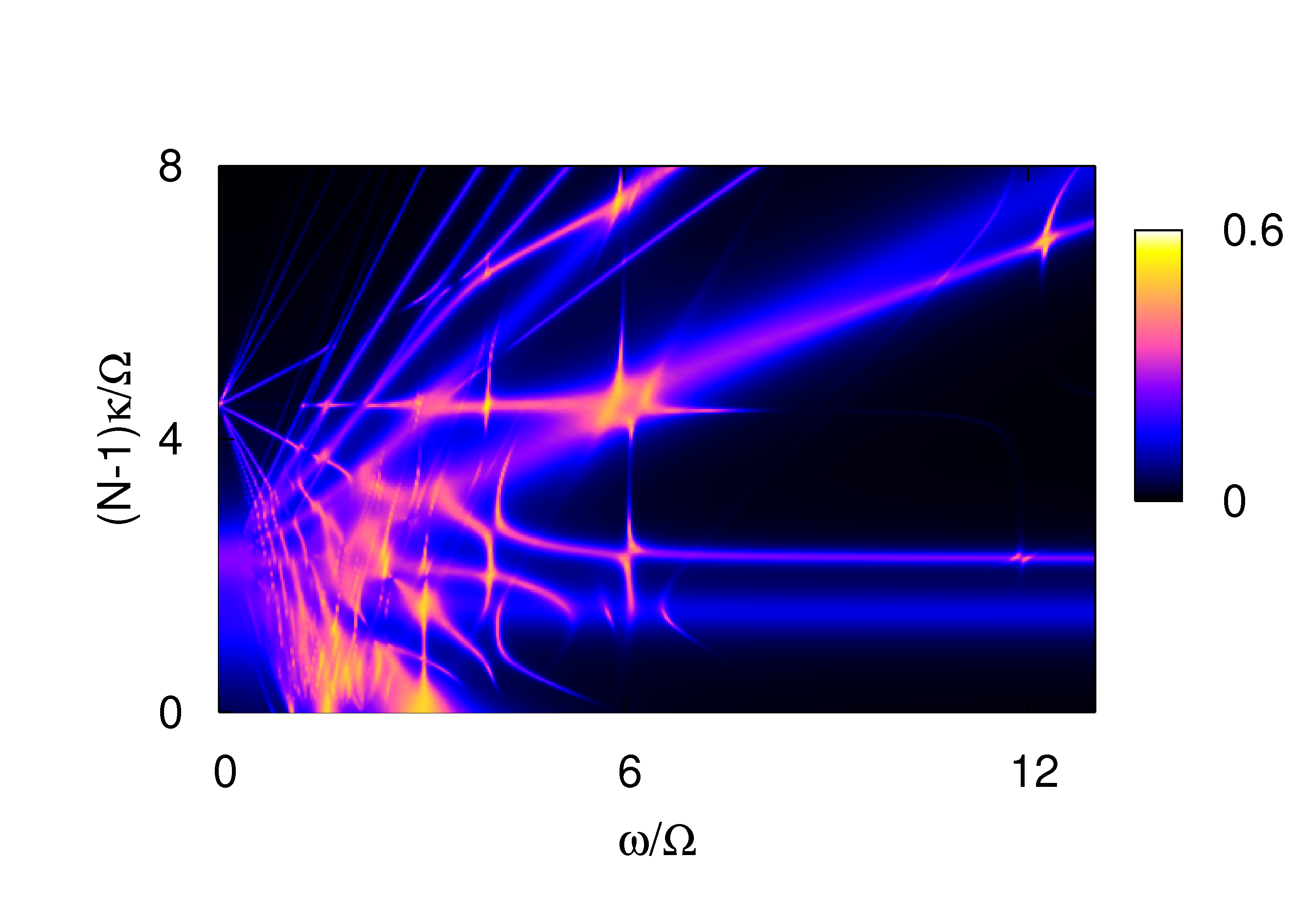}
\caption{\label{fig:4_100}(Colour online) Time-averaged transfer calculated for the same parameters as Fig.~\ref{fig:4_10} except for $\Omega T =100$. While the 1/2-photon resonance is clearly visible near $\omega = 6\Omega$, the 1/4-resonance can also be found for some interactions near  $\omega = 12\Omega$. }
\end{figure}

More interesting features, including fractional photon resonances, emerge for larger averaging times (Fig.~\ref{fig:4_100}). While the short-time effects visible in Fig.~\ref{fig:4_10} can be explained by simply looking at tunnelling of a single particle, Fig.~\ref{fig:4_100} displays many features which are due to many-particle tunnelling. For larger particle numbers, similar features will only be visible on even larger time-scales (cf.\ Sec.~\ref{sec:N}). In Fig.~\ref{fig:4_100}, both the 1/2-photon and the 1/4-photon resonance are visible for a broad range of interaction strengths.

Integer-photon resonances essentially are single-particle effects which survive interactions. Fractional photon assisted tunnelling, however, is a true many-particle quantum effect. For four particles, one expects~\cite{XieEtAl10} to observe the 1/2- and the 1/4-resonance both of which are clearly visible in Fig.~\ref{fig:4_100}. Odd fractions like the 1/3-photon resonance (cf.\ Sec.~\ref{sec:N}) only occur for odd particle numbers $N\ge 3$ (cf.\ Ref.~\cite{XieEtAl10}).
Fractional resonances also appear for the case of $N=100$ (cf.\ Ref.~\cite{EckardtEtAl05}). However, for the experimentally motivated comparatively short timescales used both here and in Ref.~\cite{EckardtEtAl05}, fractional photon-assited tunnelling is a very small effect even for small BECs. We therefore focus on the case of few-atoms per well for the remainder of the paper.

\section{\label{sec:4}The 1/2- and 1/4- resonance}
For fractional resonances in a double well lattice with few atoms per double well, most of the physics of the tunnelling process can be understood by taking the approach of Sec.~\ref{sec:understanding} to a level beyond perturbation theory. Within the approximation motivated by the rotating wave-approximation, this leads to time-dependent $(N+1)\times (N+1)$ matrices which can, in some cases, even be solved analytically (see the appendix). Figure~\ref{fig:onehalf} shows that the 1/2-photon resonance will be clearly visible for four particles. As the full width at half maximum is an order of magnitude larger than 1\% (the typical error~\cite{Trotzky10} with which the interaction can be fixed in experiments as Ref.~\cite{CheinetEtAl08}), the 1/2-photon resonance could be observed with the existing experimental setup of Refs.~\cite{CheinetEtAl08,ChenEtAl11}.
\begin{figure}
\includegraphics[angle=-90,width=\linewidth]{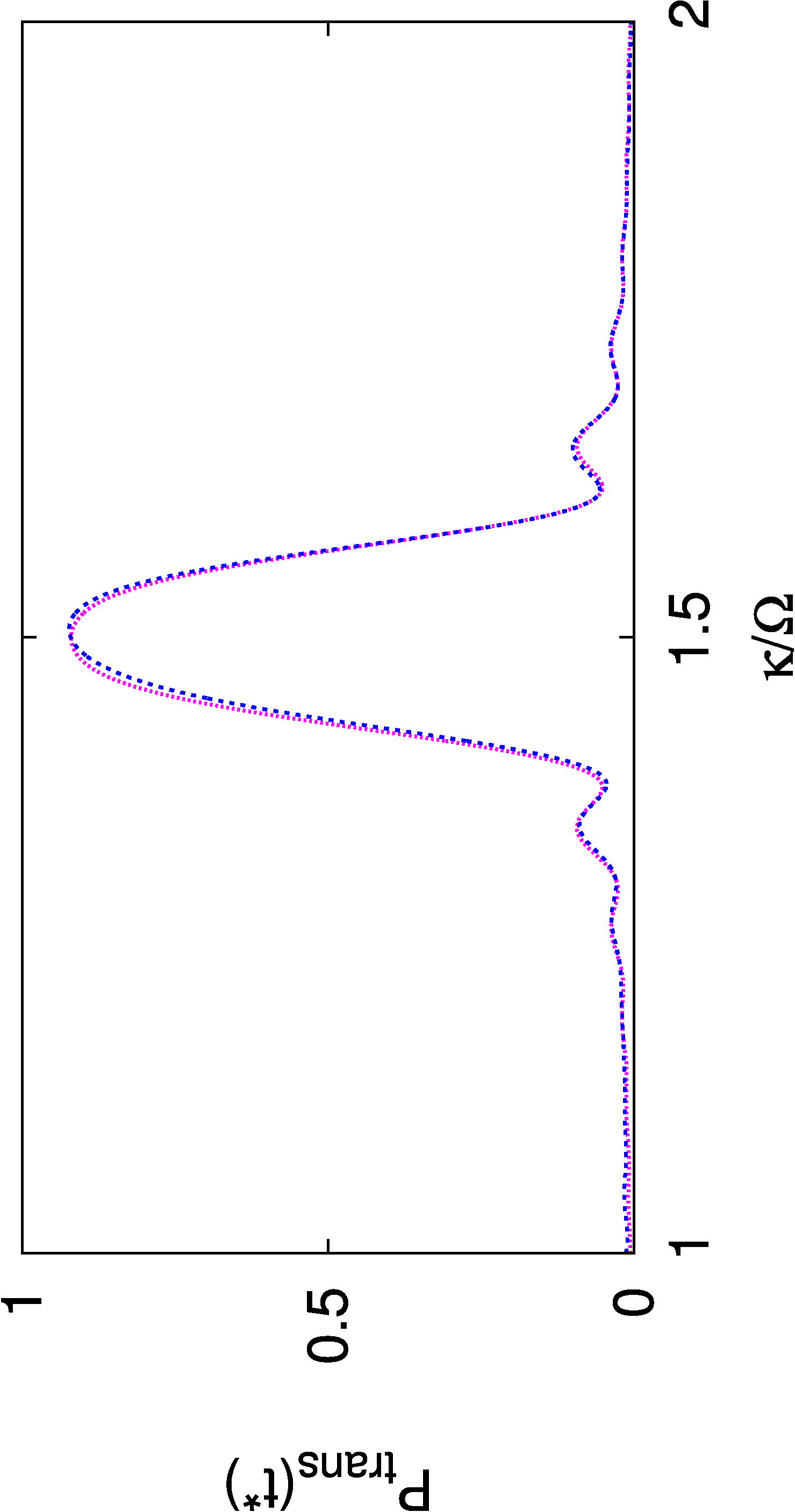}
\caption{\label{fig:onehalf}(Colour online) 1/2-photon resonance for four particles; displayed is the transfer to the second well at time $t^*=13.115$ as a function of the scaled interaction parameter $\kappa/\Omega$. Dotted/magenta: model discussed in the appendix [Eq.~(\ref{eq:figonehalf})], dashed/blue: full numerics. The parameters are: $\omega = 6\Omega$, $\mu_0=1.5\Omega$, $2\mu_1/\Omega=4.567$. As the full width at half maximum is an order of magnitude larger than the typical experimental accuracy for $\kappa/\Omega$~\cite{Trotzky10},  the 1/2-photon resonance could be observed with existing experimental setups. Choosing the correct time will also be feasible: For $\kappa=1.5\Omega$, numerics shows that missing the time $t^*$ by 10\% only leads to deviations of $P_{\rm trans}(t)$  from $P_{\rm trans}(t^*)\simeq 0.92$ by less than 3\%. }
\end{figure}

\begin{figure}
\includegraphics[width=\linewidth]{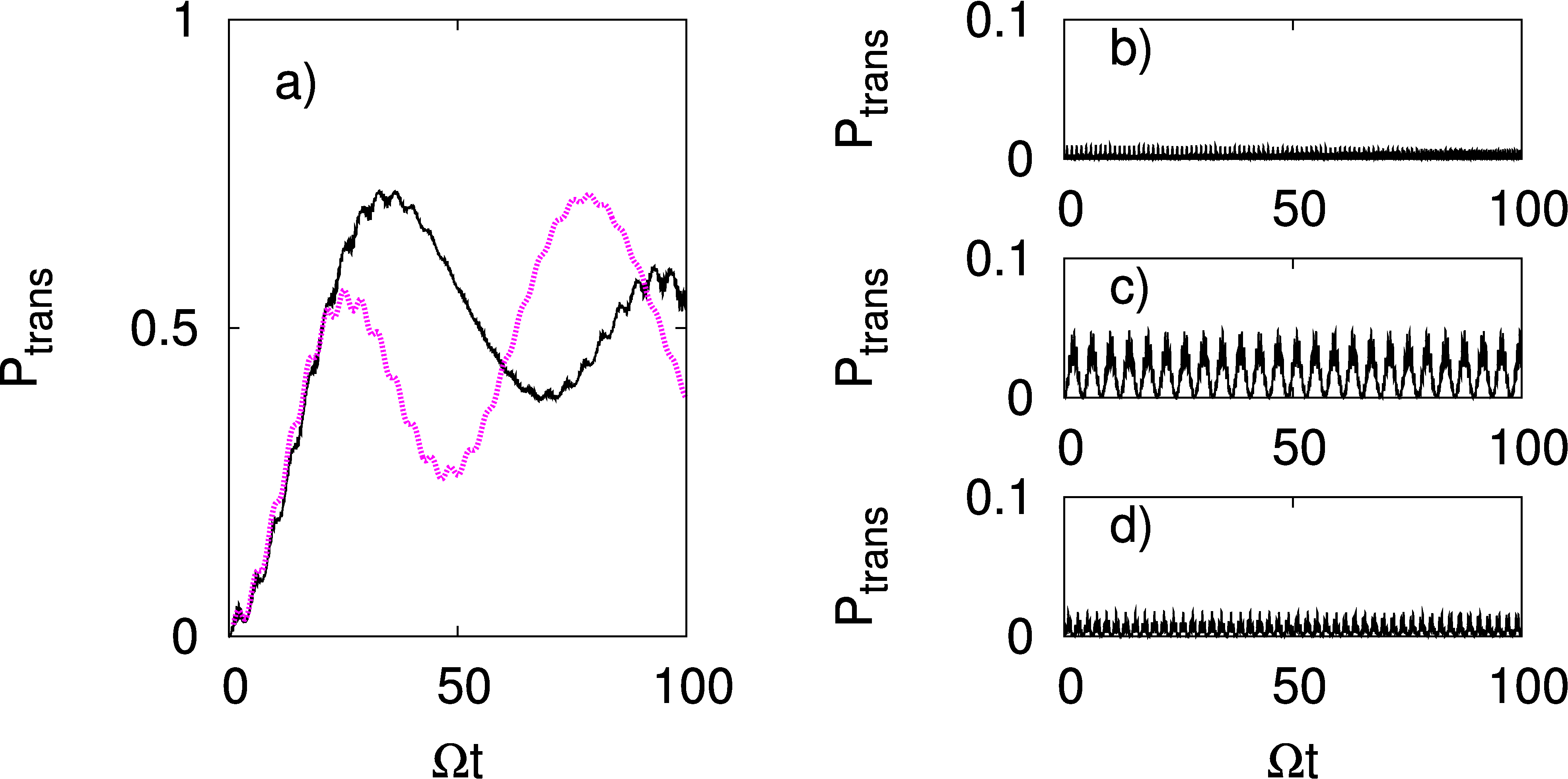}
\caption{\label{fig:onequarter}(Colour online) a) 1/4-photon resonance for four particles ($\kappa = 2.25\Omega$, $\omega = 12\Omega$; $2\mu_1/\omega = 3.08$, $\mu_0= 1.5\Omega$). Black line: exact numerics. The simplified model (magenta/grey line) correctly describes the dynamics for not too large times. In order to demonstrate that this is really an effect which only occurs for four particles, $N<4$ was also investigated: b) N=3, c) N=2, d) N=1. Note that the scale on the vertical axis differs by an order of magnitude between a) and b)-d).}
\end{figure}

Given the fact that the 1/2-photon resonance has already been observed experimentally~\cite{ChenEtAl11}, it would be even more interesting to investigate the 1/4-photon resonance as $N=4$ is the lowest number of particles for which it can be observed. Just because 4 particles produce a large effect in the numerics does, however, not automatically imply that it is a true 4-particle effect: Many features which are already visible for the short averaging times of Fig.~\ref{fig:4_10} would also occur at least similarly for lower particles. Thus, even if we restrict our search for a 1/4-resonance to parameters near $\omega=12\Omega$, this could coincide with large tunnelling for lower particle numbers. Due to the experimental way to load double-well lattices via a Mott-insulator~\cite{ChenEtAl11}, the harmonic confinement will prevent experiments with all double-wells being filled with exactly 4 particles. However, as can be clearly seen from Fig.~\ref{fig:onequarter}, only the wells initially loaded with 4 atoms can contribute to the observable signature of the 1/4-photon resonance. 

\section{\label{sec:avoided}Avoided-crossing-type features}
\begin{figure}[t]
\includegraphics[angle=-90,width=\linewidth]{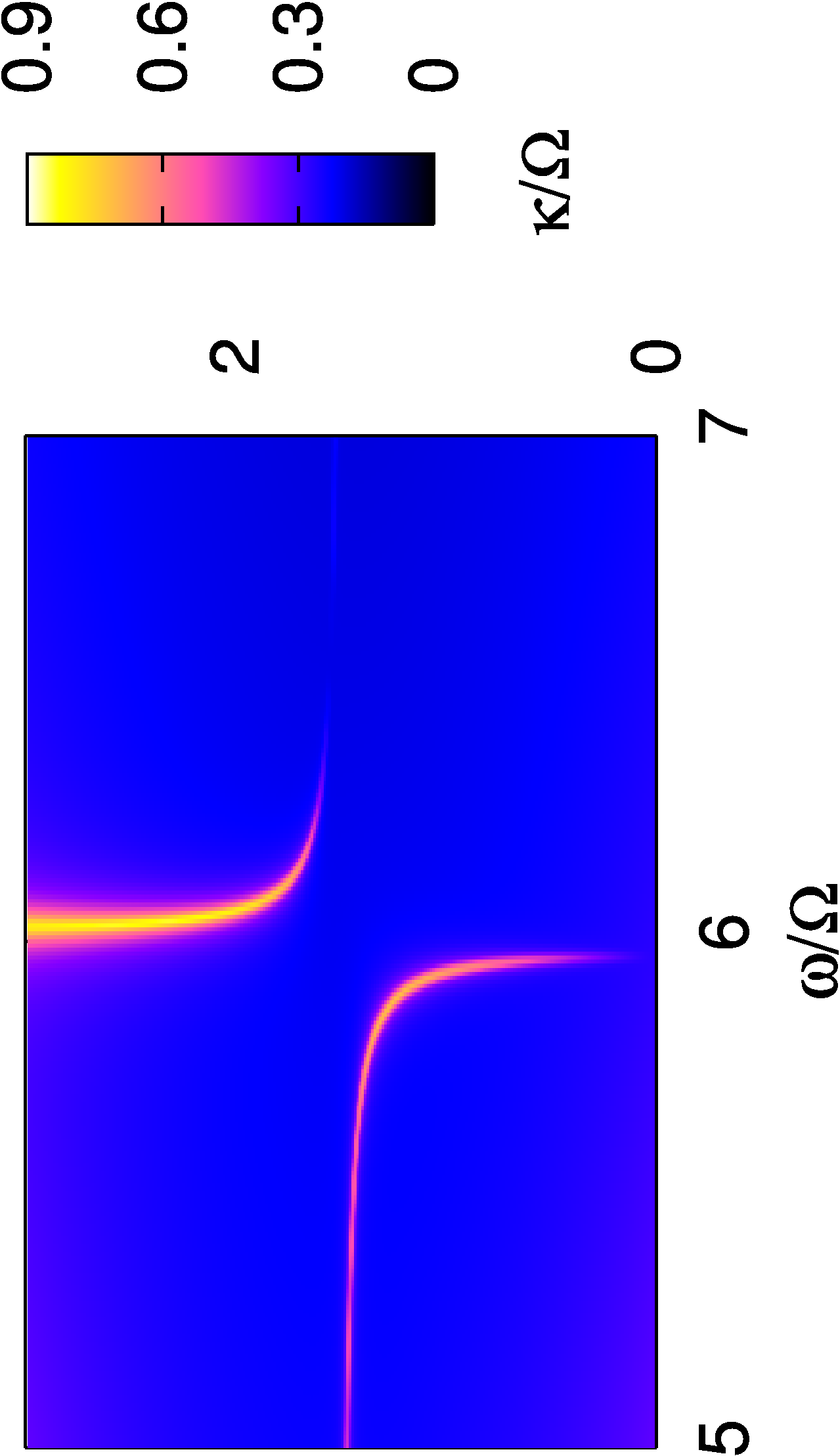}
\caption{\label{fig:avoided}(Colour online) Two-dimensional projection of the fourth root of the time-averaged transfer for $N=2$ particles. The averaging time is $T\Omega=500$, the tilt is $\mu_0=1.5\Omega$ and the shaking amplitude ${2\mu_1}/{\omega}=5.5201$ (for which $|J_0\left({2\mu_1}/{\omega}\right)|<10^{-5}$). In Sec.~\ref{sec:avoided}, these parameters allow the explanation of an example how the apparent avoided-crossing-type features (cf.\ Fig.~\ref{fig:4_100}) can appear. }
\end{figure}
Figure~\ref{fig:4_100} shows several features which resemble avoided crossings. The smallest particle number for which they can occur is $N=2$ for which the feature is particularly strong at the 1/2-photon frequency near zeros of the Bessel function responsible for the tunnelling of the first particle. Such a situation is depicted in Fig.~\ref{fig:avoided}.  For parameters near the centre of this figure, i.e.\ $\mu_0 = 1.5\Omega$, $\omega =6\Omega$ and $\kappa=1.5\Omega$, the tunnelling of the first particle would normally be described by a 0-photon process while the tunnelling of the second particle would be a 1-photon process. Only the average number of photons per tunnelling process justifies the word ``1/2-photon resonance''. If, however, $J_0\left({2\mu_1}/{\omega}\right)=0$ (as chosen for Fig.~\ref{fig:avoided}), we have an entirely different situation. Now the tunnelling of the first particle consists of two competing processes: a $1$-photon process and a $-1$-photon process, making the overall tunnelling a superposition of a 0-photon and a 2-photon process [see Eq.~(\ref{eq:nphoton})]. As $J_{-1}(x)=-J_{1}(x)$, we have:
\begin{align}
\label{eq:avoid_a1}
a_1^{(1)} =& \frac{\Omega}{\sqrt{2}}J_1\left({\textstyle\frac{2\mu_1}{\omega}}\right)\left[\frac{\exp\left(i\eta_1^{(1)}t\right)}{i\eta_1^{(1)}}+\frac{\exp\left(i\eta_{-1}^{(1)}t\right)}{i\eta_{-1}^{(1)}}\right]\nonumber \\
-&\frac{\Omega}{\sqrt{2}}J_1\left({\textstyle\frac{2\mu_1}{\omega}}\right)\left[\frac{1}{i\eta_1^{(1)}}+\frac{1}{i\eta_{-1}^{(1)}}\right]
\end{align}
This equation already explains why there is no transfer visible at $\mu_0 = 1.5\Omega$, $\omega =6\Omega$ and $\kappa=1.5\omega$, as 
\begin{equation}
\eta_1^{(1)} =-\eta_{-1}^{(1)}
\end{equation}
for these parameters. In the vicinity of this point, the time-independent part of Eq.~(\ref{eq:avoid_a1}) will be responsible for $a_2$ eventually becoming larger, in particular, for parameters for which $\eta^{(2)}_1=0$:
\begin{align}
\label{eq:avoid_a2}
a_2^{(2)} \sim& \frac{\Omega^2}{2}J^2_1\left({\textstyle\frac{2\mu_1}{\omega}}\right)\left[\frac{1}{i\eta_1^{(1)}}+\frac{1}{i\eta_{-1}^{(1)}}\right]t\;.
\end{align}
Combined with the fact that $\eta^{(2)}_1=0$ corresponds to the line 
\begin{equation}
\kappa=\frac{\omega-3\Omega}2,
\end{equation}
this explains why transfer becomes large when following
this line away from the point ($\omega=6\Omega$, $\kappa=1.5\Omega$); for the line perpendicular to this
line through the same point such increased transfer is neither to be expected nor does it show in the numerics of Fig.~\ref{fig:avoided}.

\section{\label{sec:N}$1/N$-resonance: large effect only in $N$th order perturbation theory}

The system investigated here offers the unique possibility to construct
physically relevant examples for which the perturbation theory is small up to $(N-1)$th order while the $N$th order produces results that dominate the dynamics.

If 
\begin{equation}
\eta_{k_j}^{(j)}\neq 0 
\end{equation}
for all $j=1\ldots N$ and furthermore\footnote{NB: The above conditions are necessary to construct examples for which $N$th order perturbation theory is the first to be large. It is, however, possible to find parameters for which the $1/N$-resonances do not meet all those conditions (cf.\ Fig.~\ref{fig:avoided}). Furthermore, $\left|a_N^{(N)}\right|$ being large is, in general, neither a necessary nor a sufficient condition for $\left|a_N\right|$ to be large.}
\begin{equation}
\sum_{\nu=\ell}^j\eta_{k_{\nu}}^{(\nu)}
\left\{\begin{array}{lcc}
\neq 0 &:& \ell < j<N \\
=0 &:& \ell=1,\;j=N
\end{array} 
\right.,
\end{equation}
perturbation theory will give only (small) oscillatory terms up to $(N-1)$th order. However, in $N$th order we obtain a term linear in time proportional to:
 \begin{align}
\label{eq:avoidleading}
\left(a_N^{(N)}(t)\right)_{\rm leading} \propto& \left(\prod_{j=1}^NJ_{k_j}\left({\textstyle\frac{2\mu_1}{\omega}}\right)\right)t\;.
\end{align}
However, just because Eq.~(\ref{eq:avoidleading}) becomes large this does not automatically imply that this can be observed in the numerics: as soon as $a_N^{(N)}(t)$ becomes large,  $a_{N-1}(t)$ changes which in turn influences  $a_{N}(t)$. Nevertheless, there are examples for which Eq.~(\ref{eq:avoidleading}) well describes the tunnelling:

\begin{figure}
\includegraphics[width=\linewidth]{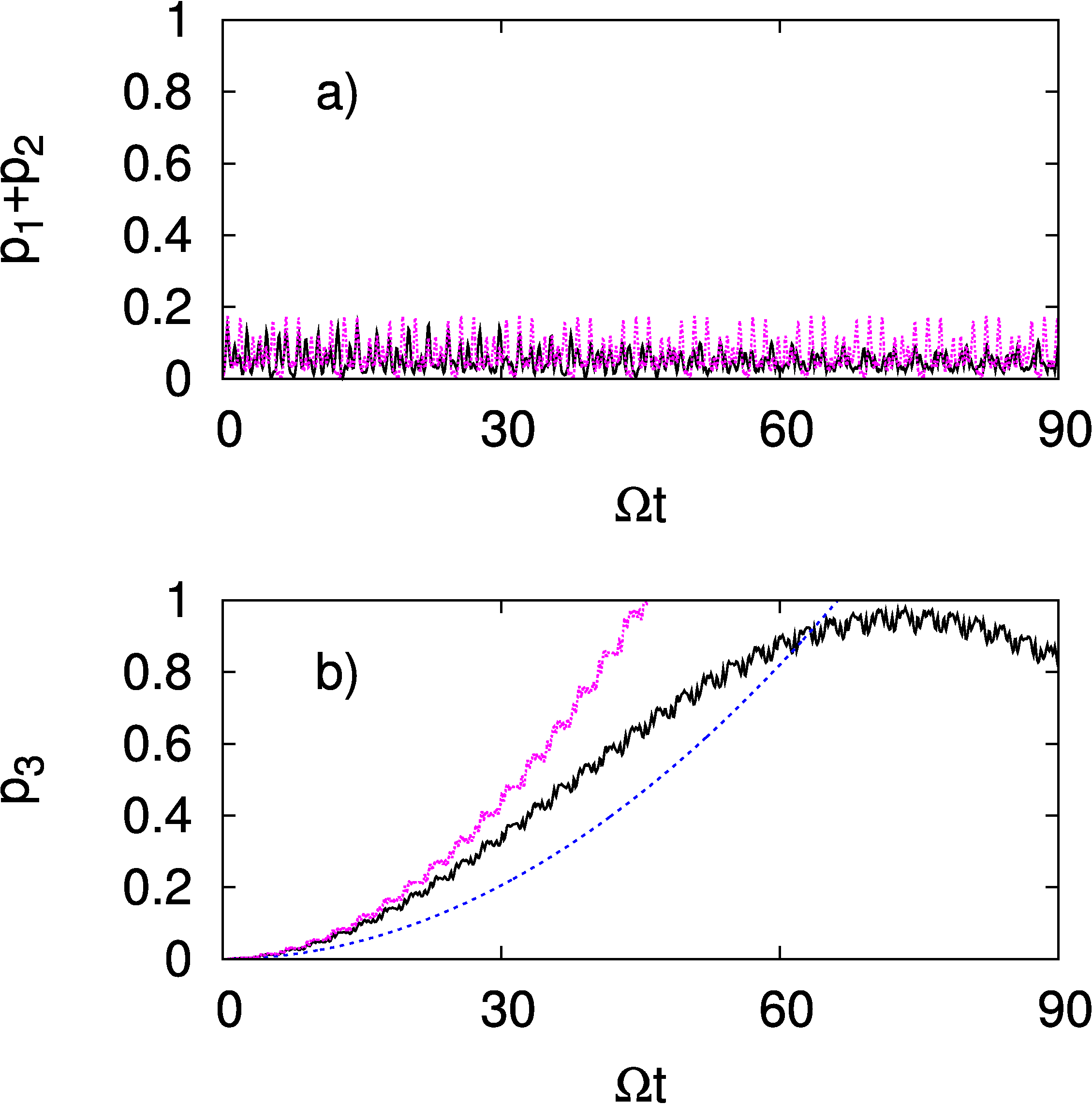}
\caption{\label{fig:N3} (Colour online) Probability to find $n$ particles in the upper well if initially all $N=3$ particles are in the lower well for $\mu_0=1.5\Omega$, $\kappa=2.1\Omega$, $\omega=9\Omega$ and $2\mu_1/\omega =0.75$. a) Black solid line: numerics using the model~(\ref{eq:H}); magenta/grey dotted line: as predicted by numerically evaluated perturbation theory by adding $|a_1^{(1)}|^2+|a_2^{(2)}|^2$ as defined in Eq.~(\ref{eq:leading}). The sum is much smaller than the probability to transfer all 3 particles which is displayed in the lower panel. b) Solid black line:  numerics using the model~(\ref{eq:H}), dotted magenta/grey line: $|a_3^{(3)}|^2$ as defined in Eq.~(\ref{eq:leading}), dashed blue/black line: analytic estimate of Eq.~(\ref{eq:leadingN3}).}
\end{figure}

For $N=3$, the above criteria are fulfilled, e.g., for $\mu_0=1.5\Omega$, $\kappa=2.1\Omega$, $\omega=9\Omega$ and $2\mu_1/\omega=0.75$:
\begin{align}
\eta_0^{(1)} &= -5.4\Omega\\
\eta_0^{(2)} &= 3\Omega\\
\eta_1^{(3)} &= 2.4\Omega.
\end{align}
 The tunnelling can thus be understood to be a 1-photon-process, the photon being responsible for the tunnelling of the third particle with no photons being involved in the tunnelling of the other two. Thus:
\begin{align}
\label{eq:leadingN3}
\left(a_3^{(3)}(t)\right)_{\rm leading} = {\textstyle\frac{3\Omega^3}{4\eta_0^{(1)}\left(\eta_0^{(1)}+\eta_0^{(2)}\right)}}J_{0}^2\left({\textstyle \frac{2\mu_1}{\omega}}\right)J_{1}\left({\textstyle \frac{2\mu_1}{\omega}}\right)\,t
\end{align}

Figure \ref{fig:N3} shows that Eq.~(\ref{eq:leadingN3})  well describes the time-scale on which the probability for all particle having tunnelled to the other well becomes large. The figure furthermore shows that the probability to find either one or two particles in the upper well remains small and this effect can thus be labelled co-tunnelling of three particles. Equation~(\ref{eq:leadingN3}) does, however, not precisely match the result of the numerical perturbation theory of Sec.~\ref{sec:timedependent}. Higher order processes would change the prefactors in both equations. For $N=3$ there is, e.g., the $+1$-, $-1$-, $+1$-photon process which also contributes to the linear time-dependence, although it is low compared with the dominating $0$-, $0$-, $+1$-photon process included in the analytic result. Note that changing the driving amplitude might change which is the dominating contribution.

\begin{figure}
\includegraphics[width=\linewidth]{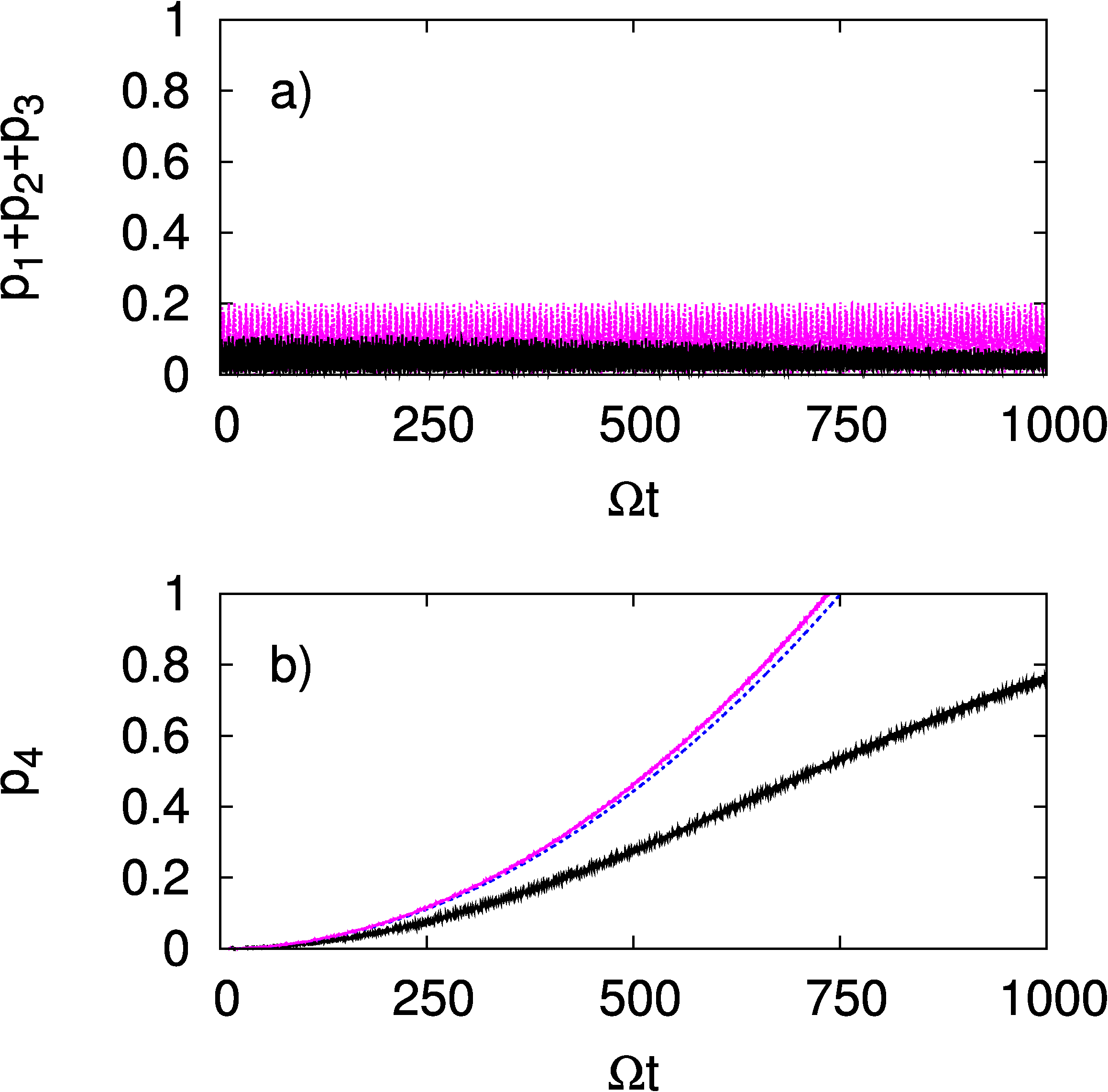}
\caption{\label{fig:N4} Probability to find $n$ particles in the upper well if initially all $N=4$ particles are in the lower well for $\mu_0=1.5\Omega$, $\omega=12\Omega$ $\kappa=1.95\Omega$, and $2\mu_1/\omega=0.5$. a) Black line: numerics using the model~(\ref{eq:H}); magenta/grey line: as predicted by numerically evaluated perturbation theory by adding $|a_1^{(1)}|^2+|a_2^{(2)}|^2+|a_3^{(3)}|^2$ as defined in Eq.~(\ref{eq:leading}). The sum is much smaller than the probability to transfer all 3 particles which is displayed in the lower panel. b) From bottom to top:  numerics using the model~(\ref{eq:H}), analytic estimate of Eq.~(\ref{eq:leadingN4}) , numeric evaluation of $|a_4^{(4)}|^2$ as defined in Eq.~(\ref{eq:leading}). The upper two curves lie very close together.}
\end{figure}

For $N=4$ similar parameters can be found. Using $\mu_0=1.5\Omega$, $\omega=12\Omega$, $\kappa=1.95\Omega$, and $2\mu_1/\omega=0.5$, one has as the leading order contribution:
\begin{align}
\eta_0^{(1)} &= -8.7\Omega\\
\eta_0^{(2)} &= -0.9\Omega\\
\eta_0^{(3)} &= 6.9\Omega\\
\eta_1^{(4)} &= 2.7\Omega,
\end{align}
with 
\begin{equation}
\label{eq:leadingN4}
\left|\left(a_4^{(4)}(t)\right)_{\rm leading}\right| = {\textstyle
\frac{3\Omega^4J_{0}^3\left({\textstyle \frac{2\mu_1}{\omega}}\right)J_{1}\left({\textstyle \frac{2\mu_1}{\omega}}\right)}{2\left|\eta_{0}^{(1)}\left(\eta_{0}^{(1)}+\eta_0^{(2)}\right)\left(\eta_{0}^{(1)}+\eta_0^{(2)}+\eta_0^{(3)}\right)\right|}}\,t\;.
\end{equation}
This analytic function correctly predicts that the tunnelling of all 4 particles at once takes place at a much longer time-scale than for three particles (Fig.~\ref{fig:N4}). As for the previous figure, the tunnelling can be labelled a co-tunnelling process, now of four particles.


\section{Conclusion}

Both perturbation-theory and beyond-perturbation-theory approaches can be applied to understand the physics of fractional photon-assisted tunnelling in optical super-lattices. In some cases even analytical calculations are possible. While the 1/2-resonance has already been observed experimentally, we predict that both the 1/3-photon resonance and the 1/4-photon resonance should be observable experimentally.

For the 1/3-resonance, an interesting phenomenon occurs: although both first and second order perturbation theory are small, third order perturbation theory nevertheless correctly describes it is a large effect. This effect can also be found for larger particle numbers. It will, however, be much more realistic to try and observe this effect for 3 particles than for much larger particle numbers: Both the time-scales for the transfer to the upper well becomes large and the model~(\ref{eq:H}) becomes less valid.

In the same way in which it is possible to describe the 1/2-resonance for some parameters  alternatively as co-tunnelling of a pair of atoms~\cite{ChenEtAl11}, we thus numerically observe co-tunnelling for both $3$ and $4$ atoms.

\acknowledgments

ME thanks E.\ Demler for the hospitality in his group. CW thanks S.\ A.\ Gardiner and M.\ Holthaus for their support. We thank Y.\ Chen, S.\ F\"olling, T.\ Monteiro, N.\ Teichmann and S.\ Trotzky for discussions. ME acknowledges funding by both the \textit{Studienstiftung des deutschen Volkes} and  the \textit{Floyd und Lili Biava Stiftung}.

\begin{appendix}
\section{Understanding the tunnelling for $N=4$ via beyond-perturbation theory calculations}
Reference~\cite{EsmannEtAl11} demonstrates that it is possible, at least for cases with one dominant frequency for each tunnelling event, to use a beyond-perturbation theory version of Sec.~\ref{sec:understanding}. In the present Section, the analytic calculations within the effective model~\cite{EsmannEtAl11} are performed in detail for $N=4$ particles. 

Using the abbreviations
\begin{align}
\label{eq:defOmega}
\Omega_k^{(j)}&\equiv  \frac{\sqrt{j}\sqrt{N-j+1}}{{2}}\Omega J_k(2\mu_1/\omega)\\
B_k^{(j)} &=  i^k\Omega_k^{(j)},
\end{align}
we can rewrite the equations of Sec.~\ref{sec:understanding} in such a way that they can be solved beyond the perturbation theory approach. We are, however, restricted to cases where only one frequency plays a role. For those we have:
\begin{equation}
\frac{d}{dt}\left(\begin{array}{c} {a}_0(t)\\ a_1(t)\\a_2(t)\\a_3(t)\\a_4(t)\end{array}\right)= \textbf{A}(t) \left(\begin{array}{c} a_0(t)\\ a_1(t)\\a_2(t)\\a_3(t)\\a_3(t)\end{array}\right)
\end{equation}
where $\textbf{A}(t)$ is a Hermitian $5\times5$ matrix which has only non-zero entries for $(\textbf{A}(t))_{n,m}$ with $n=m\pm1$: 
\begin{equation}
 \label{eq:A1}
(\textbf{A}(t))_{n,n+1}= B_{k_n}^{(n)}\exp\left(-i\eta_{k_n}^{(n)}t\right)
\end{equation}
and
\begin{equation}
 \label{eq:A2}
(\textbf{A}(t))_{n+1,n}= (B_{k_n}^{(n)})^*\exp\left(i\eta_{k_n}^{(n)}t\right).
\end{equation}
This can be solved using the ansatz
\begin{equation}
\label{eq:ansatz}
\left(\begin{array}{c} {a}_0(t)\\ a_1(t)\\a_2(t)\\a_3(t)\\a_4(t)\end{array}\right)
= 
\left(\begin{array}{c} \widetilde{a}_0\exp(-i\omega t)\\ \widetilde{a}_1\exp[-i(\omega-\eta_{k_1}^{(1)})t]\\\widetilde{a}_2\exp[-i(\omega-\eta_{k_1}^{(1)}-\eta_{k_2}^{(2)})t]\\\widetilde{a}_3\exp[-i(\omega-\eta_{k_1}^{(1)}-\eta_{k_2}^{(2)}-\eta_{k_3}^{(3)})t]\\\widetilde{a}_4\exp[-i(\omega-\eta_{k_1}^{(1)}-\eta_{k_2}^{(2)}-\eta_{k_3}^{(3)}-\eta_{k_4}^{(4)})t]\end{array}\right)
\end{equation}
which leaves us to calculate the eigenvalues of the time-independent matrix:
\begin{equation}
 \label{eq:BN4}
\textbf{B}=
\left(\begin{array}{ccccc}
0 & -B_{k_1}^{(1)} & 0 & 0 & 0\\
-B_{k_1}^{*(1)} & \eta_{k_1}^{(1)} & -B_{k_2}^{(2)} & 0 & 0\\
0 & -B_{k_2}^{*(2)} & \eta_{k_1}^{(1)}+\eta_{k_2}^{(2)} & -B_{k_3}^{(3)} & 0\\
0 & 0 & -B_{k_3}^{*(3)} & \sum_{j=1}^3 \eta_{k_j}^{(j)} & -B_{k_4}^{(4)}\\
0 & 0 & 0 & -B_{k_4}^{*(4)} & \sum_{j=1}^4 \eta_{k_j}^{(j)}\\
\end{array}\right)\;.
\end{equation}
In order to obtain a $1/2$-photon resonance the following conditions~\cite{TeichmannEtAl09} are imposed on the choice of the ideal frequencies $\eta_{k_j}^{(j)}$:
\begin{eqnarray}
\eta_{k_1}^{(1)}+\eta_{k_4}^{(4)}=0 &\Rightarrow& k_1+k_4=1\;,\nonumber\\
\eta_{k_2}^{(2)}+\eta_{k_3}^{(3)}=0 &\Rightarrow& k_2+k_3=1\;,\nonumber\\
\eta_{k_1}^{(1)}+\eta_{k_2}^{(2)}=0 &\Rightarrow& \kappa=-\frac{1}{8}(k_1+k_2)\omega-\frac{1}{2}\mu_0\;.
\end{eqnarray}
Furthermore $\omega/2=2\mu_0$ and thus one may choose $\kappa=\omega/4$ implying $k_1+k_2=-1$. Once choosing $k_1$ all frequencies are fixed. With the above conditions fulfilled it is now possible to solve the eigenvalue problem analytically. For the sake of readability from now on the following further changes in the notation will be adopted in the four-particle case:
\begin{eqnarray}
\label{eq:simple}
\Omega_{k_1}^{(1)} &\rightarrow& \Omega_1\;,\nonumber\\
\Omega_{-k_1-1}^{(2)} &\rightarrow& \Omega_2\;,\nonumber\\
\Omega_{k_1+2}^{(3)} &\rightarrow& \Omega_3\;,\nonumber\\
\Omega_{1-k_1}^{(4)} &\rightarrow& \Omega_4\;,\nonumber\\
\eta_{k_1}^{(1)} &\rightarrow& \eta_1\;.
\end{eqnarray}
The eigenvalues $\omega_i$ are:
\begin{eqnarray}
\label{eq:eigenvalues}
\omega_1 &=& 0\;,\nonumber\\
\omega_{2/3} &=& \frac{1}{2}\left\{\eta_1\mp\sqrt{\eta_1^2+2\left(\Omega_1^2+\Omega_2^2+\Omega_3^2+\Omega_4^2-\sqrt{K}\right)}\right\}\;,\nonumber\\
\omega_{4/5} &=& \frac{1}{2}\left\{\eta_1\mp\sqrt{\eta_1^2+2\left(\Omega_1^2+\Omega_2^2+\Omega_3^2+\Omega_4^2+\sqrt{K}\right)}\right\}\;,\nonumber\\
\end{eqnarray}
where $K$ represents the expression:
\begin{eqnarray}
\label{eq:K}
K &\equiv& \Omega_1^4+\Omega_2^4+2\Omega_2^2(\Omega_3^2-\Omega_4^2)+\nonumber\\
&+& 2\Omega_1^2(\Omega_2^2-\Omega_3^2-\Omega_4^2)+(\Omega_3^2+\Omega_4^2)^2\;.
\end{eqnarray}
The corresponding eigenvectors $\widetilde{v}^{(i)}$ are inserted into the ansatz~(\ref{eq:ansatz}) and yield five independent, not yet normalised, solutions to the time-dependent problem:
\begin{eqnarray}
\label{eq:eigenvectors}
v^{(1)}(t) &=& \left(\begin{array}{c}
-\frac{\Omega_2\Omega_4}{\Omega_1\Omega_3}\\
0\\
i\frac{\Omega_4}{\Omega_3}\\
0\\
1
\end{array}\right)\;,\nonumber\\
v^{(2)}(t) &=& \left(\begin{array}{c}
\frac{\Omega_1(-\Omega_1^2-\Omega_2^2+\Omega_3^2+\Omega_4^2+\sqrt{K})}{2\Omega_2\Omega_3\Omega_4}\\
-\frac{i^{-k'}(-\Omega_1^2-\Omega_2^2+\Omega_3^2+\Omega_4^2+\sqrt{K})}{2\Omega_2\Omega_3\Omega_4}\omega_2e^{i\eta_1t}\\
-\frac{i(\Omega_1^2+\Omega_2^2+\Omega_3^2-\Omega_4^2-\sqrt{K})}{2\Omega_3\Omega_4}\\
-\frac{i^{-k'+1}}{\Omega_4}\omega_2e^{i\eta_1t}\\
1
\end{array}\right)e^{-i\omega_2t}\;,\nonumber\\
v^{(3)}(t) &=& \left(\begin{array}{c}
\frac{\Omega_1(-\Omega_1^2-\Omega_2^2+\Omega_3^2+\Omega_4^2+\sqrt{K})}{2\Omega_2\Omega_3\Omega_4}\\
-\frac{i^{-k'}(-\Omega_1^2-\Omega_2^2+\Omega_3^2+\Omega_4^2+\sqrt{K})}{2\Omega_2\Omega_3\Omega_4}\omega_3e^{i\eta_1t}\\
-\frac{i(\Omega_1^2+\Omega_2^2+\Omega_3^2-\Omega_4^2-\sqrt{K})}{2\Omega_3\Omega_4}\\
-\frac{i^{-k'+1}}{\Omega_4}\omega_3e^{i\eta_1t}\\
1
\end{array}\right)e^{-i\omega_3t}\;,\nonumber\\
v^{(4)}(t) &=& \left(\begin{array}{c}
-\frac{\Omega_1(\Omega_1^2+\Omega_2^2-\Omega_3^2-\Omega_4^2+\sqrt{K})}{2\Omega_2\Omega_3\Omega_4}\\
\frac{i^{-k'}(\Omega_1^2+\Omega_2^2-\Omega_3^2-\Omega_4^2+\sqrt{K})}{2\Omega_2\Omega_3\Omega_4}\omega_4e^{i\eta_1t}\\
-\frac{i(\Omega_1^2+\Omega_2^2+\Omega_3^2-\Omega_4^2+\sqrt{K})}{2\Omega_3\Omega_4}\\
-\frac{i^{-k'+1}}{\Omega_4}\omega_4e^{i\eta_1t}\\
1
\end{array}\right)e^{-i\omega_4t}\;,\nonumber\\
v^{(5)}(t) &=& \left(\begin{array}{c}
-\frac{\Omega_1(\Omega_1^2+\Omega_2^2-\Omega_3^2-\Omega_4^2+\sqrt{K})}{2\Omega_2\Omega_3\Omega_4}\\
\frac{i^{-k'}(\Omega_1^2+\Omega_2^2-\Omega_3^2-\Omega_4^2+\sqrt{K})}{2\Omega_2\Omega_3\Omega_4}\omega_5e^{i\eta_1t}\\
-\frac{i(\Omega_1^2+\Omega_2^2+\Omega_3^2-\Omega_4^2+\sqrt{K})}{2\Omega_3\Omega_4}\\
-\frac{i^{-k'+1}}{\Omega_4}\omega_5e^{i\eta_1t}\\
1
\end{array}\right)e^{-i\omega_5t}\;.\nonumber\\
\end{eqnarray}
Any state $\left|\Psi(t)\right\rangle$ of the four-particle system can now be expressed as a linear combination  $\left|\Psi(t)\right\rangle=b_1v^{(1)}(t)+b_2v^{(2)}(t)+b_3v^{(3)}(t)+b_4v^{(4)}+b_5v^{(5)}$ of the eigenstates. The coefficients $b_i$ are again obtained from the initial condition $\left|\Psi(0)\right\rangle\equiv\left|0\right\rangle$. 

The probability to find the system in a Fock state $\left|i\right\rangle$ is given by the overlap with $\left|\Psi(t)\right\rangle$ and may be expressed as the square of an amplitude $\left|a_i(t)\right|^2$ with $v_i^{(j)}$ being the $i$-th component of the $j$-th eigenvector:
\begin{equation}
\left|\left\langle \Psi(t)\left|i\right.\right\rangle\right|=\left|a_i(t)\right|^2=\left|\sum_{j=1}^5b_j\frac{v_{i+1}^{(j)}(t)}{\left|v^{(j)}(t)\right|}\right|^2\;.
\end{equation}
Finally, the time-dependent transfer $P_{\rm trans}(t)$ is expressed in terms of the amplitudes $a_i(t)$ as
\begin{equation}
\label{eq:figonehalf}
P_{\rm trans}(t)=\frac{1}{4}\left[4\left|a_4(t)\right|^2+3\left|a_3(t)\right|^2+2\left|a_2(t)\right|^2+\left|a_1(t)\right|^2\right]\;.
\end{equation}
This expression for given time and amplitude of the driving is displayed as a function of the coupling strength in Fig.~\ref{fig:onehalf}.

\end{appendix}


\end{document}